\documentstyle[twocolumn,aps]{revtex}             
\input epsf

\begin{document}
\title{Evidence for a singularity in ideal magnetohydrodynamics:
implications for fast reconnection {\rm $\quad\quad$ (Submitted to \prl)}}
\author{Robert M. Kerr$^1$ and Axel Brandenburg$^2$}
\address{$^1$NCAR, Boulder, CO 80307-3000;
$^2$Mathematics, University of Newcastle, NE1 7RU, UK}

\maketitle

\begin{abstract}
Numerical evidence for a finite-time singularity in ideal 3D
magnetohydrodynamics (MHD) is presented. 
The simulations start from two interlocking magnetic flux
rings with no initial velocity.  Curvature shrinks the rings
until they touch and current sheets form between them.
The evidence for a singularity in a finite time $t_c$ is that
the peak current density behaves like $\|J\|_\infty\sim 1/(t_c-t)$
for a range of sound speeds 
and initial conditions.
For the incompressible calculations 
$\|\mbox{\boldmath$\omega$}\|_\infty/\|{\bf J}\|_\infty\rightarrow\mbox{const}$.
In resistive reconnection the magnetic helicity is nearly
conserved while energy is dissipated.
\end{abstract}
\pacs{PACS number: 52.30.Jb, 52.65.Kj}

To date it is not known whether or not the ideal magnetohydrodynamic (MHD)
and Euler equations are regular. Regularity means that for arbitrary
smooth initial data the velocity ${\bf u}$ and/or 
magnetic field ${\bf B}$ and all
of its derivatives remain finite for all times. If regularity is not
established, one cannot rule out the possibility of a finite time singularity,
which for ideal MHD implies satisfying a new mathematical constraint that 
the vorticity $\mbox{\boldmath$\omega$}=\mbox{\boldmath$\nabla$}\times{\bf u}$
and current density ${\bf J}=\mbox{\boldmath$\nabla$}\times{\bf B}$
must obey \cite{CKS97}
\begin{equation}
I(t_c)=\int_0^{t_c} 
\left[\|\mbox{\boldmath$\omega$}\|_\infty(t)+\|{\bf J}\|_\infty(t)\right] dt 
\rightarrow\infty,
\label{eq:CKS}
\end{equation}
for there to be a singularity.
Singularities associated with shocks are also possible, but are not
considered here, because they would not occur in the incompressible limit.
The physical significance of a finite-time singularity is that fast
vortex and/or magnetic field reconnection could be possible once
viscous and resistive effects are restored.  A theoretical demonstration
of a mechanism for fast magnetic reconnection would be significant
in a variety of problems in plasma physics including 
experimental studies related to magnetically
confined fusion \cite{PPPL97-98}, dynamos, the earth's magnetic field
and the solar corona, where 
reconnection is associated with flares \cite{OhyamaShibata97,Yohkoh92}
and coronal heating \cite{Parker94,GalsgaardN96}.
This letter will show preliminary evidence for 
a possible singularity for ideal MHD using numerical simulations
of three-dimensional linked magnetic flux rings
and emphasize its relationship to fast magnetic reconnection.

Significant progress has been made recently for the Euler case where
a numerical study of a pair of antiparallel vortex
tubes has produced strong evidence for the formation of a singularity
in the Euler equations in a finite time \cite{Kerr93}. The numerical
evidence was consistent with an analytic constraint for Euler
\cite{BKM84} that no
singularity can occur in a finite time $t=t_c$ 
in any quantity (e.g. in any derivative of
${\bf u}$, however high) unless
\begin{equation}
I(t_c)=\int_0^{t_c} \|\mbox{\boldmath$\omega$}\|_\infty(t) dt \rightarrow\infty.
\label{eq:BKM}
\end{equation}
Here $\|...\|_\infty$ is the $L^\infty$ norm, or maximum, in space.
This theorem  shows that when searching
for a singularity in the 3D Euler equations, the only quantity that needs
to be monitored numerically is $\|\mbox{\boldmath$\omega$}\|_\infty$. 
Furthermore, if a singularity of the form 
$\|\mbox{\boldmath$\omega$}\|_{\infty}  \sim
(t_c-t)^{-\gamma}$ is observed in a numerical experiment 
then $\gamma$ must obey
$\gamma\geq 1$ for the observed singular behavior to be genuine.
The numerical comparisons \cite{Kerr93} found
$\gamma\equiv 1$, a scaling that is consistent dimensionally in addition
to being consistent with (\ref{eq:BKM}).  
The generalization to MHD is (\ref{eq:CKS}).

An important feature of the analysis of the Euler calculations 
that should be used when analysing other flows with possible singularities
is that singular behavior should be demonstrated
by several independent tests.
One set of tests predicted mathematically is
that in addition to the $1/(t_c-t)^\gamma$ behavior of 
$\|\mbox{\boldmath$\omega$}\|_\infty$,
all components of $\|\mbox{\boldmath$\nabla$}{\bf u}\|_\infty$ should diverge 
as $1/(t_c-t)^\gamma$ \cite{Ponce85}. Again, $\gamma\equiv1$ is expected.
Another test that was found numerically \cite{Kerr93} is that
the rate of enstrophy production is
\begin{equation}
d\Omega/dt=\int \omega_i e_{ij} \omega_j\,dV \sim 1/(t_c-t)
\label{eq:enstp}
\end{equation}
where enstrophy is
${\textstyle{1\over2}}\Omega=\int |\mbox{\boldmath$\omega$}|^2\,dV$.  

The question of fast magnetic reconnection has been addressed using
both time evolving calculations and steady-state analyses. In
two-dimensional calculations, fast reconnection is inhibited by
material that cannot escape easily from between thin current sheets. 
This led to early suggestions that a singularity in 2D MHD
is precluded by the formation of current sheets \cite{Pouquet80s} and
is consistent with a recent result that there
can be no singularity at 2D nulls \cite{Klapper2D}. 
New experimental evidence \cite{PPPL97-98} 
designed to test two-dimensional steady-state theories 
shows a slow resistive timescale \cite{Sweet56,Parker57,Parker63}
instead of a faster timescale resulting from slow shocks \cite{Petschek64}
and related theories \cite{PriestForbes92}.  
In three dimensions, the extra degree of freedom could allow material
to escape more readily from the current sheets.  This has prompted
reconnection simulations in three dimensions, for example
starting with orthogonal, straight flux tubes in pressure equilibrium
and a small velocity field to push them together
\cite{DahlburgAntiochos95}. Explosive growth of $\|{\bf J}\|$ has
been seen before \cite{Batyetal98}, but not in the context of singularities.

We consider the equations for an isothermal, compressible gas 
for a given sound speed $c_s$ with a magnetic field of the form
$$
{\partial{\bf u}\over\partial t}
=-{\bf u}\cdot\mbox{\boldmath$\nabla$}{\bf u}
-c_s^2\mbox{\boldmath$\nabla$}\ln\rho
+{{\bf J}\times{\bf B}\over\rho}+{\mu\over\rho}\left(\nabla^2{\bf u}
+{\textstyle{1\over3}}\mbox{\boldmath$\nabla$}
\mbox{\boldmath$\nabla$}\cdot{\bf u}\right),
$$
\begin{equation}
{\partial\ln\rho\over\partial t}
=-{\bf u}\cdot\mbox{\boldmath$\nabla$}\ln\rho
-\mbox{\boldmath$\nabla$}\cdot{\bf u},
\label{eq:isoth}
\end{equation}
$$
{\partial{\bf A}\over\partial t}={\bf u}\times{\bf B}+\eta\nabla^2{\bf A},
$$
where ${\bf B}=\mbox{\boldmath$\nabla$}\times{\bf A}$ is the magnetic field 
in terms of the magnetic vector potential ${\bf A}$, 
${\bf u}$ is the velocity, and $\rho$ is the density.
In the ideal limit, the resistivity
$\eta$ and the viscosity $\mu$ are set to zero.
The magnetic field is measured in units where the permeability is unity. 
Periodic boundary conditions are adopted in
a domain of size $(2\pi)^3$. Our time unit is the sound travel time over
a unit distance. The equations are advanced in time
using a variable third-order Runge-Kutta timestep 
and sixth order explicit centered finite differences in space. 

The equations for the incompressible case are the same except that the equation
for $\rho$ is replaced by the divergence-free condition on
velocity $\mbox{\boldmath$\nabla$}\cdot{\bf u}$ to determine the pressure.
The magnetic field ${\bf B}$, rather than the vector potential ${\bf A}$,
is used as a fundamental variable.
The equations are advanced in time using a spectral colocation method with 
the 2/3-rule and variable third-order Runge-Kutta timestep.
The maximum resolution used $384^3$ mesh points.

The incompressible, ideal MHD equations conserve total energy
$E={1\over2}\int({\bf u}^2+{\bf B}^2)\,dV$,
magnetic helicity $H_B=\int{\bf A}\cdot{\bf B}\,dV$ and the cross helicity
$H_C=\int{\bf u}\cdot{\bf B}\,dV$.
The helicities can be used to describe aspects of 
the topology \cite{Moffatt69}.  
Since $H_B$ has one less spatial
derivative than the energy, spectrally it 
should dissipate more slowly than energy \cite{Berger84} when
$\mu,\nu\neq0$.

The initial conditions used in the present study will
all be of two linked, magnetic flux rings.
This condition is chosen because it has the advantage that
no velocity field needs 
to be imposed in the initial conditions since the
tension from the curvature of the rings induces a velocity
by shrinking the rings. Due to the initial linkage, 
it also yields nearly maximal $H_B$ and therefore is
an excellent choice for studying the dissipation of $H_B$
versus energy \cite{Moffatt69}.  

\begin{figure}[t]
\epsfxsize=8.5cm
\epsfbox{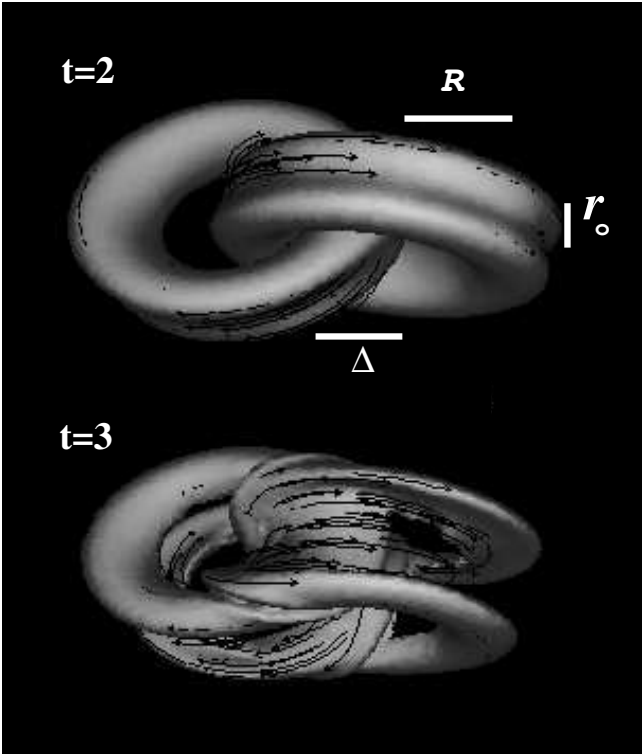}
\caption[]{Resistive calculation using compressible code.
$t=2$ shows evolution during the nearly ideal phase, $t=3$ shows a
partially reconnected state with $H_B$ converted more into
new twist between the remnants of the original tubes than
into writhe within reconnected tubes.}
\label{fig:writhe}\end{figure}

A variety of different ring thicknesses and angles between
the initial rings have been
investigated.  This letter will discuss in detail only cases
where the rings are orthogonal and are just touching.  
There are three distances that determine the initial condition
that evolved into the structures in Figure~\ref{fig:writhe}: the
radii of the rings $R$, the thickness $r_\circ$
where the flux goes smoothly to zero, and the
separation of their centers from the origin $\Delta$.  
The separation of the rings is $2(R-\Delta)$.
An initial profile across the ring that gives $|B|=1$ in the center
and goes smoothly to $|B|=0$ at $r=r_\circ$ is
taken from the Euler calculations \cite{Kerr93}.  
$R=1$ and $\Delta=0.5$ for all the cases, 
so that they go through each other's centers.
The initial condition for the compressible calculations to be reported used 
$r_\circ=0.5$.  $r_\circ=0.5$, 0.65 and 0.8
for the ideal incompressible cases.
Following the example from the Euler case \cite{Kerr93},
the following hyperviscous filter was applied
in Fourier space to the initial condition only:
$\exp[-(k/k_{\max})^4)$, where $k_{\max}$ is between 14 and 20.
As a result, the maximum initial magnetic field $B_0$ is slightly
less than one. 
For the compressible calculations,
$c_s$ is varied between 0.1 and 10, so the initial plasma beta,
$\beta_0=2c_s^2/B_0^2$, varied between 0.02 and 200. Only nearly
incompressible $\beta_0=2$ calculations are presented.  The initial
density was uniform and unity. While what is being presented is
nearly incompressible, preliminary low plasma beta calculations
as well as compressible vortex reconnection \cite{VirkHK95}
suggest that compressibility might enhance reconnection rates
and singularity formation.

All of the compressible, resistive calculations were 
straightforward runs from $t=0$
with viscosities and resistivities chosen for a given resolution. 
The strategy for reaching the highest
possible resolution for the ideal calculations followed the
example of how a possible singularity
in incompressible Euler was demonstrated \cite{Kerr93}. 
First, the ideal calculations
do not contain any numerical smoothing and have been run only
so long as numerical dissipation was insignificant.  This approach
was used because experience has shown that artificial
smoothing results in artificial dissipation which can obscure
the dynamics of the ideal case.  The lower resolution calculations 
were started from $t=0$.  
Since it would be too expensive to
run higher resolution calculations from $t=0$, 
they are being remeshed at intermediate times.  

There was no initial velocity field.
The first, and shortest, phase after initialization was that due
to the curvature of the flux rings their diameter $R$ shrinks.
This automatically brings the flux tube rings into contact and the current
of one ring begins to overlap the magnetic field of the other.  This
is necessary for the Lorentz force to be significant and for a strong
interaction to begin. 

\begin{figure}[t]
\epsfxsize=8.5cm
\epsfxsize=8.5cm
\epsfbox{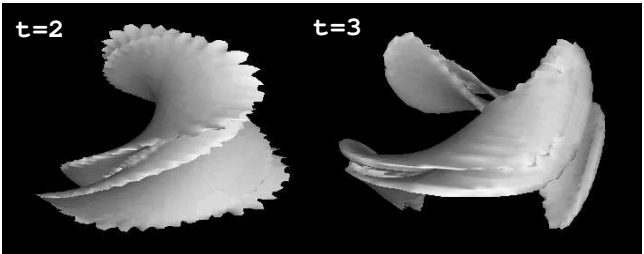}
\caption[]{Three-dimensional level surfaces of the magnitude of the current
density in the resistive calculation. $t=2$ shows a double saddle surface, which
at $t=3$ is broken up into two disjoint pieces.}
\label{fig:current}\end{figure}

To give an overall view of the flow, Figure~\ref{fig:writhe} shows
the three-dimensional structure from a $\beta=0.5$,
$192^3$ resistive, compressible calculation,
just before and after the estimated singular time.
$t=2$ shows nearly ideal evolution from the
initial motionless, perfectly tubular flux tubes.  The dominant
feature is the indentation in each ring.
This is because in the cores of the flux rings the field is
strongest and the magnetic curvature force largest.
The region going singular appears in Figure~\ref{fig:current}a.
as twisted current sheets within the center of
Figure~\ref{fig:writhe}a.
A surprising property of the reconnection process is that 
slices show that the size of the entire structure shrinks.  
Some perspective on this
can be obtained by comparing the structures at $t=2$ and $t=3$.

By $t=3$ some reconnection has occurred.  Magnetic helicity $H_B$
is nearly conserved, decreasing linearly at a very slow pace as
it is converted into writhe \cite{MR92} or new twist,
and energy is dissipated more rapidly.  
A detailed examination of this structure will be studied elsewhere.

In a resistive calculation, the dissipation is concentrated on a current sheet
that forms where the two flux rings come into contact.
Level surfaces of the current density near its peak value (Fig.~\ref{fig:current})
show a twisted, saddle shaped, double-sheet structure 
before the estimated singular time,
which separates into two disjoint sheets centered around the points
of maximum current density after the singular time.

\begin{figure}[t]
\epsfxsize=8.5cm
\epsfbox{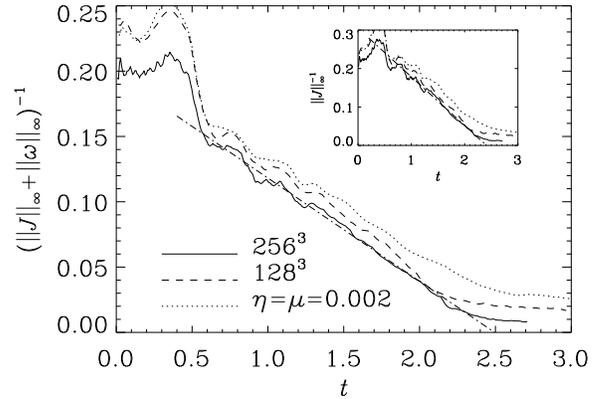}
\caption{Evolution of $1/\|{\bf J}\|_\infty$ and
$1/(\|{\bf J}\|_\infty+\|\omega\|_\infty)$. $r=0.5$ for $\beta_0=2$.
Solid and dashed lines refer to ideal calculations with different
resolution and the dotted line is for a resistive calculation with
$\eta=\mu=0.002$. The dash-dotted line gives a linear fit to the data.
}\label{fig:1Jcompr}\end{figure}

Figure~\ref{fig:1Jcompr}
shows $1/\|{\bf J}\|_\infty$ and 
$1/(\|{\bf J}\|_\infty+\|\mbox{\boldmath$\omega$}\|_\infty)$ for
the resistive calculations.
There is a strong tendency in favor of linear
behavior similar to that observed for 3D Euler \cite{Kerr93}.
Extrapolating from before $t=2.3$ to $1/J_{\max}=0$ 
suggests that $t_c\approx2.5$.
For $c_s>0.5$, that is more incompressible, roughly the same
singular time would be predicted.  For $c_s<.5$, that is more
compressible, different behavior is indicated, but the trend
toward $1/\|{\bf J}\|_\infty \sim (t_c-t)$ remains. 

For Euler, recall that stronger evidence for a possible
singularity was obtained by monitoring the $L^\infty$
norm for an additional strain term 
and the enstrophy production, as well as
$\|\mbox{\boldmath$\omega$}\|_\infty$ \cite{Kerr93}.  
Following that reasoning, we need
to know the behavior of two first derivatives of the 
magnetic or velocity fields plus a global production term. 
Therefore, we propose looking at the
behavior of $\|{\bf J}\|_\infty$,
$\|\mbox{\boldmath$\omega$}\|_\infty$, and the production of
$\Omega_{\omega+J}={\textstyle{1\over2}}
\int(|\mbox{\boldmath$\omega$}|^2+|{\bf J}|^2)\,dV$,
\begin{equation}
P_{\Omega J}=\int (\omega_i e_{ij} \omega_j 
- \omega_i d_{ij} J_j + 2\varepsilon_{ijk}J_i d_{j{\ell}}e_{{\ell}k})\,dV,
\label{eq:j2o2prod}
\end{equation} 
where 
$e_{ij}={1\over2}(u_{i,j}+u_{j,i})$ and
$d_{ij}={1\over2}(B_{i,j}+B_{j,i})$
are the hydrodynamic and magnetic strains.  
The terms in $P_{\Omega J}$, in order, 
are the vortex stretching already known for Euler, 
a new vorticity production term and a new current producing term.  
All three tests should go as $1/(t_c-t)$ once sufficiently
singular solutions are obtained, which we have not yet achieved.
Therefore, the present objective is trends in the right direction,
rather than conclusive evidence for the existence of a singularity.  

\begin{figure}[t]
\epsfxsize=8.5cm
\epsfbox{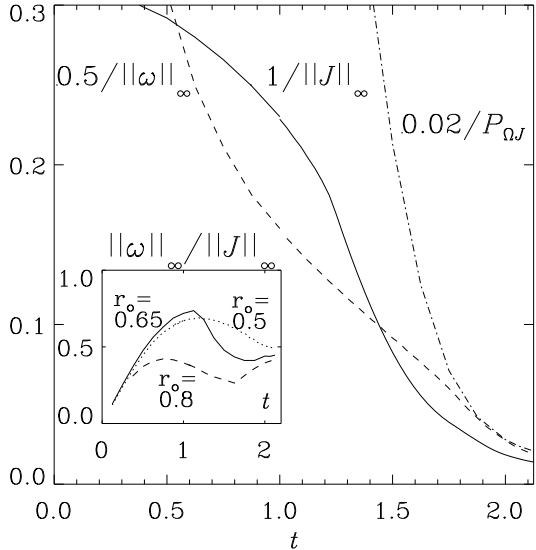}
\caption{Comparison of the evolution of $1/\|{\bf J}\|_\infty$,
$1/\|\mbox{\boldmath$\omega$}\|_\infty$ and $P_{\Omega J}$ for $r_\circ=0.65$.
The inset shows $\|\mbox{\boldmath$\omega$}\|_\infty/\|{\bf J}\|_\infty$
for $r=$0.5, 0.65 and 0.8, all of which are converging to
$\|\mbox{\boldmath$\omega$}\|_\infty/\|{\bf J}\|_\infty\approx0.5$,
demonstrating possible self-similarity.}
\label{fig:1Jincompr}\end{figure}

Figure~\ref{fig:1Jincompr} plots the three proposed tests for 
the $r_\circ=0.65$ incompressible case.  The $384^3$ run was used for
the $1/\|{\bf J}\|_\infty$ and 
$1/\|\mbox{\boldmath$\omega$}\|_\infty$ tests for $t>1$ and the
$192^3$ run for $1/P_{\Omega J}$ for all time.

In the case of Euler it was shown that ratios of all $L^\infty$ norms of
first derivatives were approaching constant values \cite{Kerr93},
which would be consistent with
self-similar behavior near the point going singular.  
Therefore we expect that 
$\|\mbox{\boldmath$\omega$}\|_\infty/\|{\bf J}\|_\infty$ should approach a
constant value here.  
Initially the velocity,
and vorticity, are zero, so these must build up before
behavior for $\|\mbox{\boldmath$\omega$}\|_\infty$ and $P_{\Omega J}$ 
consistent with $1/(t_c-t)$ appears.
The inset in Figure~\ref{fig:1Jincompr}
shows that all three cases appear to be converging to a
value of $\|\mbox{\boldmath$\omega$}\|_\infty/\|{\bf J}\|_\infty\approx0.5$.  
$r_\circ=0.65$ converges the soonest, near $t=1.6$ and together with
the convergence of 
$\|\mbox{\boldmath$\omega$}\|_\infty/\|{\bf J}\|_\infty\rightarrow0.5$ 
is our best evidence to date for a possible singularity in ideal MHD.
In addition, $1/P_{\Omega J}$ appears to join the same $(t_c-t)$ near $t=1.9$.

While $1/\|{\bf J}\|_\infty$ for the $r_\circ=0.5$ incompressible
case is comparable to the compressible case and shows the largest range of 
$1/\|{\bf J}\|_\infty\sim(t_c-t)$ behavior, it is not strong evidence
for a singularity because the inset in Figure~\ref{fig:1Jincompr} shows that 
its $1/\|\mbox{\boldmath$\omega$}\|_\infty$
up to $t=2.0$ does not have the same value for $t_c$.
However, note that the production term for ${\bf J}^2$ in 
(\ref{eq:j2o2prod}), $2\varepsilon_{ijk}J_i d_{j{\ell}}e_{{\ell}k}$,
does not involve vorticity, but only strain terms.  
Unlike Euler, vorticity seems to
be playing only a secondary role for ideal MHD.
It is not essential
for vorticity to blow up for the compressible cases, and it does not.
Related to this, the current sheets near $\|{\bf J}\|_\infty$ 
in Figure~\ref{fig:current} are twisted. 
This also suggests that similar hydrodynamic initial conditions should
be revisited to determine their reconnection rates \cite{ArefZ91}.
Further analysis of local production terms in
proposed $600^3$ calculations should help answer these questions.

In conclusion, we have presented numerical evidence for a finite-time
blow-up of the current density in ideal MHD in the case of interlocked magnetic
flux rings. In the resistive case one would not expect there to be a
singularity. Instead, arbitrarily thin current sheets will form, depending
on how small the resistivity is \cite{Otto95}.
This can lead to significant dissipation
whose strength is virtually independent of resistivity. 
The astrophysical significance of current
sheets as a consequence of tangential discontinuities of the field has
also been stressed in a recent book by Parker \cite{Parker94}. However,
the possibility of a finite time singularity in ideal MHD is not commonly 
discussed in connection with fast reconnection. 

This work has been supported in part by an EPSRC visiting grant GR/M46136.
NCAR is support by the National Science Foundation.  We appreciate
suggestions by J.D. Gibbon, I. Klapper, and H.K. Moffatt.

\end{document}